\documentclass[
reprint,
superscriptaddress, 
amsmath
,amssymb,
aip, 
apl,
]{revtex4-1}

\usepackage{graphicx}
\usepackage{dcolumn}
\usepackage{changes}
\usepackage[textsize=tiny]{todonotes}
\usepackage[utf8]{inputenc}
\usepackage[scientific-notation=false,alsoload=synchem]{siunitx}
\usepackage{soul}

\newcommand{\aGadSet}{$\alpha$-Ga$_2$Se$_3$\ }
\newcommand{\GadSet}{Ga$_2$Se$_3$\ }
\newcommand{\bGadOt}{$\beta$-Ga$_2$O$_3$\ }

\newcommand{\aSe}{$a$-Se\ }
\newcommand{\Fref}[1]{Fig. \ref{#1}}
\newcommand{\fref}[1]{Fig. \ref{#1}}

\begin{document}
\setlength{\belowdisplayskip}{1pt} \setlength{\belowdisplayshortskip}{1pt}
\setlength{\abovedisplayskip}{1pt} \setlength{\abovedisplayshortskip}{1pt}

\title{Oxidation dynamics of ultrathin GaSe probed through Raman spectroscopy}

\author{Alaric Bergeron}
\email{alaric.bergeron@polymtl.ca}
\affiliation{Department of Engineering Physics, Polytechnique Montréal, Montréal H3T~1J4, Canada}

\author{John Ibrahim}
\affiliation{Department of Engineering Physics, Polytechnique Montréal, Montréal H3T~1J4, Canada}

\author{Richard Leonelli}
\affiliation{Department of Physics, Université de Montréal, Montréal H3T~1J4, Canada}

\author{Sebastien Francoeur}
\email{sebastien.francoeur@polymtl.ca}
\affiliation{Department of Engineering Physics, Polytechnique Montréal, Montréal H3T~1J4, Canada}


\begin{abstract}
Gallium selenide (GaSe) is a 2D material with a thickness-dependent gap, strong non-linear optical coefficients and uncommon interband optical selection rules, making it interesting for optoelectronic and spintronic applications. In this work, we monitor the oxidation dynamics of GaSe with thicknesses ranging from 10 to 200 nm using Raman spectroscopy. In ambient temperature and humidity conditions, the intensity of all Raman modes and the luminescence decrease rapidly with moderate exposure to above-gap illumination. Concurrently, several oxidation products appear in the Raman spectra: Ga$_2$Se$_3$, Ga$_2$O$_3$ and amorphous and crystalline selenium. We find that no safe measurement power exists for optical measurements on ultrathin GaSe in ambient conditions. We demonstrate that the simultaneous presence of oxygen, humidity, and above-gap illumination is required to activate this photo-oxidation process, which is attributed to the transfer of photo-generated charge carriers towards aqueous oxygen at the sample surface, generating highly reactive superoxide anions that rapidly degrade the sample and quench the optical response of the material.
\end{abstract}

\maketitle

Gallium selenide (GaSe) is a layered pseudo-direct gap semiconductor\cite{Nagel1979} with strong non-linear properties in the visible, IR and THZ ranges   \cite{Allakhverdiev2009,Mandal2008}, a high photoresponse from 2 to 5 eV  \cite{Ho2006b}, a highly anisotropic band structure exhibiting uncommon selection rules \cite{Liang1975,Kuroda1981}, and interesting spin physics  \cite{Starukhin2010}. Atomically thin GaSe can be mechanically exfoliated from bulk samples  \cite{Late2012,Hu2012} or grown by chemical vapor transport  \cite{Lei2013}. Its bulk band gap of \SI{2.12}{\electronvolt}  \cite{Aulich1969a,Allakhverdiev2009} is predicted to increase to more than \SI{3.5}{\electronvolt} for the monolayer  \cite{Rybkovskiy2011,Hu2013}. Atomically thin flakes and devices have shown a high photoresponse \cite{Li2014d,Hu2012}, a composition tunable band gap  \cite{Jung2015}, near-unity optical polarization  \cite{Tang2015}, the strongest second-harmonic generation observed in a monolayer 2D material  \cite{Karvonen2015, Zhou2015a}, and transistors with high ON/OFF ratios \cite{Late2012b}. 

Like most 2D materials, the properties of GaSe are expected to sensitively depend on the interaction of its surface with its chemical environment. In bulk form, GaSe is generally considered as a stable material and is known to have a high laser damage threshold suitable for non-linear optics applications \cite{Fernelius1994,Allakhverdiev2013}. Nonetheless, bulk GaSe is also known to naturally form a native oxide \cite{Drapak2008a} and thermally- and photo-induced oxidation has been reported  \cite{Katerynchuk2014, Iwakuro1982, Kovalyuk2008}. Recently, exposure to intense laser light was found to degrade optical properties and lead to chemical transformations \cite{Beechem2015,Andres-Penares2017,Susoma2017}. Given the interest for few-layer GaSe for electronic and optoelectronic applications, it appears critical to examine the oxidation dynamics of ultrathin GaSe and determine the conditions in which this oxidation can be suppressed or at least minimized.

In this paper, we monitor the complex oxidation dynamics of ultrathin GaSe and identify the emergence of several oxidation products using Raman spectroscopy. This oxidation, highly detrimental to all optical and electronic processes, requires the simultaneous presence of oxygen, humidity, and above-gap illumination and is adequately described in the framework of the Marcus-Gerischer theory. We demonstrate that removing any of these three components greatly reduces the oxidation rate. 

GaSe layers were mechanically exfoliated from Bridgeman-grown bulk crystals \cite{Jandl1976} using PDMS stamps onto silicon (Si) substrates covered with a \SI{100}{\nano\meter} thermal oxide (SiO$_2$). Flakes with thicknesses ranging from 10 to \SI{200}{\nano\meter} were identified by optical contrast using an optical microscope and their thickness was established from atomic force microscopy (AFM) measurements. Raman measurements were conducted in backscattering geometry using a \SI{532}{\nano\meter} single-mode laser. The spatial and spectral resolution were \SI{0.6}{\micro\meter} and \SI{0.2}{\per\centi\meter}, respectively. For sample in vacuum experiments, samples were exfoliated in a dry N$_2$-flushed environment and transferred to an optical cell evacuated to a pressure of \SI{1.5e-6}{\milli\torr}. Deoxygenated water vapor was obtained from deionized water subjected to several freeze-thaw cycles under vacuum. Experiments in oxygen were done using 99.999\% O$_2$ gas ($<$3 ppm H$_2$O) at a partial pressure of \SI{250}{\torr}. Finally, measurements in ambient air were done at a temperature of \SI{23}{\celsius} and under a 48\% relative humidity.

\begin{figure}[htb]%
\includegraphics[width=8cm]{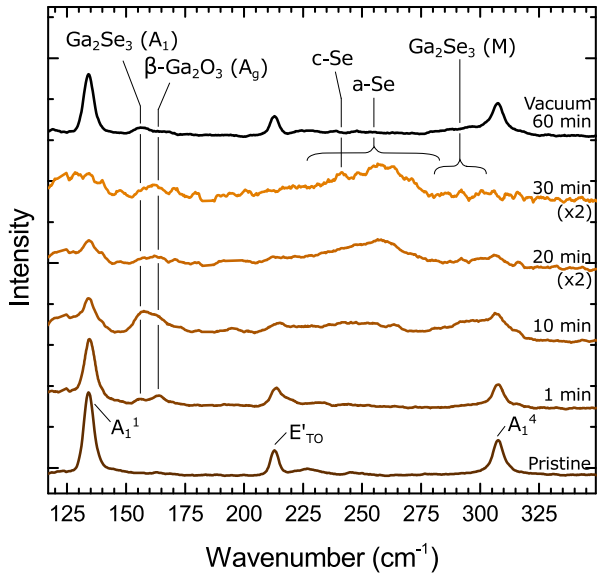}
\caption{Evolution of the Raman spectrum of a \SI{45}{\nano\meter}-thick GaSe sample in air as a function of exposure time to a  \SI{6}{\milli\watt\per\micro\meter\squared} radiation at \SI{532}{\nano\meter}. Pristine refers to a second sample measured in vacuum immediately after exfoliation; the topmost curve is a spectrum obtained after \SI{60}{\minute} of \SI{6}{\milli\watt\per\micro\meter\squared} exposure in vacuum of the same sample. In the experiment, the temperature increase induced by the exposure light is estimated at less than \SI{40}{\kelvin} }
\label{Fig:Raman}%
\end{figure}

Raman spectra are used to monitor the integrity of thin GaSe layers and the appearance of photo-induced oxidation products. In this first experiment, samples were exposed to \SI{6}{\milli\watt\per\micro\meter\squared} of radiation at \SI{532}{\nano\meter}, which corresponds to about \SI{6}{\percent} of the established laser damage threshold for nanosecond pulses at \SI{620}{\nano\meter} \cite{May1997}. This illumination is about \SI{230}{\milli\electronvolt} above the bandgap of GaSe. \Fref{Fig:Raman} presents the evolution of the Raman spectrum as a function of exposure time for a 45 nm thick GaSe sample in air. Freshly exfoliated GaSe measured in vacuum (pristine GaSe) presents the usual dominant Raman modes expected from bulk GaSe \cite{Hoff1975,Jandl1976,Wieting1972}: $A_1^1$ at {\SI{134}{\per\centi\meter}, $E_1^\prime(TO)$ at \SI{214}{\per\centi\meter}, and $A_1^4$ at \SI{308}{\per\centi\meter}}. Two other vibrational modes located at {\SI{236}{\per\centi\meter} and at \SI{246}{\per\centi\meter}} were often observed from pristine GaSe. These Raman modes are also associated to bulk GaSe  {\cite{Wieting1972,Hoff1975,Jandl1976}}, but they will not be considered in this work due to their much weaker intensity compared to $A_1^1$, $E_1^\prime(TO)$ and $A_1^4$.

The Raman spectrum evolves significantly as a function of exposure time. The intensity of all features associated with GaSe rapidly decreases and several additional Raman modes appear at frequencies of \SI{155}{\per\cm}, \SI{161}{\per\cm}, \SI{257}{\per\cm} and \SI{282}{\per\cm}. All GaSe vibrational modes are already accounted for  \cite{Hoff1975,Wieting1972} and none of these additional features can be related to any of the various known GaSe polytypes. They are attributed to a photo-induced chemical transformation of GaSe and corresponds to well-known thermal oxidation products: \GadSet, a-Se, \bGadOt (see \S SI in Supplementary Material for a description of the three main thermal oxidation pathways and their activation temperatures). 

The frequency and width of the first mode at \SI{155}{\per\cm} corresponds to the vibrational mode $A_1$ of \GadSet \cite{Finkman1975}. Although \GadSet exhibits several Raman-allowed modes, the one at \SI{155}{\per\cm} is the narrowest and most intense \cite{Yamada1992, Markl1995, Finkman1975} and thus the more likely to be observed. \GadSet has a defect zinc-blende structure in which Ga vacancies may or may not be ordered. The polarization-resolved Raman measurements presented in Fig. S1(a) in Supplementary Material suggest that the photo-induced \aGadSet phase is polycrystalline. Another feature related to \GadSet is observed at \SI{292}{\per\cm} after a 10 min photoexposure. This broad ($\sim$\SI{10}{\per\cm}) feature actually corresponds to a group of at least three vibrational modes with various symmetry representations ($A_1$, $B_1$, $B_2$) that have been assigned to localized vibrational modes of vacancy-disordered \GadSet \cite{Yamada1992}. For its connection with multiple vibrational modes, this feature is labeled \GadSet(M). Its presence suggests a high level of disorder in the \GadSet phase.

After exposure of 20 min and more, the Raman signal is dominated by two even broader features. The first at \SI{130}{\per\cm} likely corresponds to SeO$_2$ or SeO$_3$ Raman modes\cite{Anderson2000,Brassington1987}, but a quantitative analysis of this feature has proven difficult; it will not be further discussed. The other feature at \SI{257}{\per\cm} matches the characteristics of a mode associated to selenium \cite{Balitskii2002, Beechem2015} and its width ($\sim$\SI{20}{\per\cm}) suggests an amorphous phase (a-Se) \cite{Carroll1981, Baganich1991}. The narrower peak ($\sim$\SI{5}{\per\cm}) at \SI{241}{\per\cm} appears last and is associated\cite{Lucovsky1967} to a selenium crystalline phase (c-Se). As it has been often demonstrated, a-Se easily photo-crystallizes \cite{Poborchii1998, Ishida1997}. 

The feature observed at \SI{161}{\per\cm} does not correspond to any modes belonging to GaSe, \aGadSet, or Se and its oxides. However, its frequency and width match the characteristics of the $A_g$ vibrational mode of \bGadOt \cite{Onuma2014, Dohy1982}. According to Ref. \onlinecite{Dohy1982}, this feature should be the second most intense Raman feature from \bGadOt; the most intense one, found in single crystals at \SI{199}{\per\cm}, has not been unambiguously observed in this work. Polarization-resolved data is presented in Fig. S1(b) in the Supporting Information.

\begin{figure}[htb]%
\includegraphics[width=8.5cm]{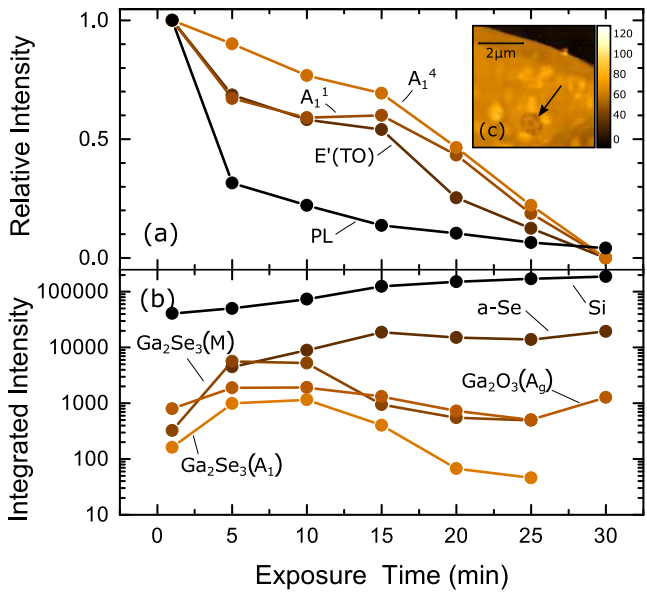}
\caption{Integrated Raman intensity and luminescence as a function of exposure time of a \SI{45}{\nano\meter} thick GaSe flake in air. (a) Relative intensity of GaSe Raman modes and photoluminescence. (b) Integrated intensity of Raman modes associated to the oxidation products and Silicon. (c) AFM surface profile (color scale in nm) of the laser damage on a \SI{58}{\nano\meter} thick flake after \SI{30}{\minute} of \SI{6}{\milli\watt} laser exposure. A black arrow indicates the $\sim$\SI{12}{\nano\meter} deep, sharply defined hole caused by the laser induced oxidation.}
\label{Fig:Oxidation}%
\end{figure}

The formation of these oxidation products is monitored as a function of time through the intensity of their Raman modes. \Fref{Fig:Oxidation} (a) shows that the relative integrated intensity of all GaSe modes ($A_1^1$, $E^\prime_{TO}$, $A_1^4$) rapidly decreases with time. After \SI{25}{\minute} of exposure, relative Raman intensities have dropped to $\sim$15\% of their initial values, revealing that the integrity of the GaSe layer is severely compromised by photo-induced oxidation mechanisms. The luminescence intensity measured at \SI{620}{\nm} also rapidly decreases (see Fig. S2 in the Supplementary Material).

\Fref{Fig:Oxidation} (b) shows the integrated intensity of the oxidation products. After only one minute of exposure, the intensity of these modes is substantial and easily measurable. \aGadSet$(A_1)$ and Ga$_2$Se$_3(M)$ increase to a maximum value after 5-\SI{10}{\minute} and then decrease below the detection limit after \SI{30}{\minute}. \bGadOt$(A_g)$ appears and then retains a relatively constant intensity throughout the oxidation process. In contrast, the intensity of a-Se rapidly increases and eventually plateaus at an intensity significantly higher than that of all other observed Raman modes, as can be seen in \fref{Fig:Raman}. 

Direct oxidation of GaSe into $\rm{Ga}_2\rm O_3$ and Se has been shown to dominate at low temperatures \cite{Balitskii2004, Drapak2008a, Stakhira2005} and, after complete oxidation involving high temperatures, only $\rm{Ga}_2\rm O_3$ and elemental selenium should remain \cite{Berchenko1997} (see \S SI of Supplementary Material). The rapid initial increase in intensity of all modes associated to Ga$_2$Se$_3$, Ga$_2$O$_3$ and Se suggests that more than one oxidation pathway is activated and that thermal oxidation is not the dominant mechanism due to the low temperatures involved. As detailed in \S II of the supplementary material, the temperature increase occurring in the experiment presented in Figs 1 and 2 is less than \SI{40}{\kelvin}.

The Raman intensity is proportional to the quantity of material in the probed volume, but it is also sensitive to its structural quality. It is therefore difficult to establish the concentrations of crystalline oxidation products, as, for example, the decrease of both Ga$_2$Se$_3$ modes could be related to its oxidation into Ga$_2$O$_3$ (see \S SI of the Supplementary Material) or to the degradation of its crystalline structure as the oxidation progresses. During oxide formation, excess Se segregates at the surface \cite{Drapak2008a}. As seen from \Fref{Fig:Oxidation}(b), its intensity quickly increases at short exposure times and mirrors the drop of intensity of GaSe Raman modes and luminescence signal. Because its response is much less sensitive to its structural quality, \aSe is believed to be the most reliable Raman marker of oxidation amongst the various products (see also \S V in Supplementary Material).

The Raman signal from the Si substrate located underneath the GaSe flake steadily increases with time even though the excitation conditions are rigorously controlled. This is explained by the generation of a large-gap oxide and localized ablation. {Oxidation leads to the formation of \bGadOt with a gap of} \SI{4.9}{\electronvolt}  \cite{Tippins1965}, which is transparent at the excitation energy of \SI{2.33}{\electronvolt}. Also, post-exposure AFM measurements indicate a layer-by-layer thinning induced by the laser exposure (see \fref{Fig:Oxidation} (c)). This is explained by the high volatility of Se and the high atomic mobility of the remaining metallic Ga \cite{Markl1995}. As a net result, the absorption from the GaSe flake decreases. 

\begin{figure}[htp]%
\includegraphics[width=8.5cm]{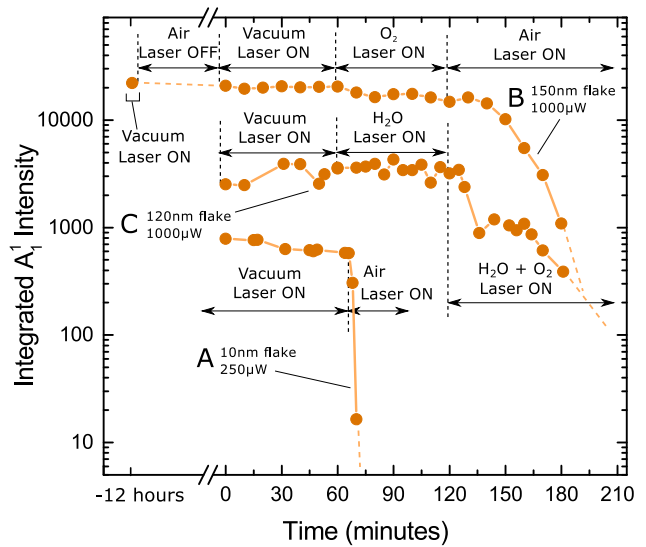}
\caption{Integrated Raman intensity of the $A_1^1$ peak of GaSe flakes as a function of exposure time to a \SI{532}{\nano\meter} laser in vacuum and in different environments. Flake thickness and exposure power are indicated on the curves.}
\label{Fig:OxidationVac}%
\end{figure}

As suggested by the data presented in \fref{Fig:Raman} (topmost curve), placing the sample in a vacuum environment protects GaSe samples from photo-induced oxidation. Curve A presented in \fref{Fig:OxidationVac} shows the evolution of the $A_1^1$ Raman intensity of a \SI{10}{\nano\meter} thick flake as a function of time for an excitation power of \SI{250}{\micro\watt}. For such thin layers, this power yields a weak but measurable Raman signal. The intensity of GaSe Raman modes is only marginally affected after an exposure of more than \SI{60}{\minute} in vacuum. At the time indicated by the vertical dashed line, the sample was exposed to ambient air. After a few minutes, the Raman intensity has dropped by two orders of magnitude and the signal is completely lost after \SI{10}{\minute}. This clearly shows that thin GaSe flakes are unstable in ambient conditions and that minimal laser illumination is sufficient to induce significant degradation, establishing that no safe above-gap excitation power exists for atomically-thin samples. 

Curve B presented in \fref{Fig:OxidationVac} shows the Raman intensity of $A_1^1$ of a \SI{150}{\nano\meter} flake exposed to \SI{1000}{\micro\watt} of laser intensity. For the first data point, the sample is in vacuum. The sample is then exposed to ambient air without any illumination. After 12 hours, the sample is again measured under vacuum conditions and the Raman intensity is very similar to that initially measured, indicating that exposure to oxygen and water vapor alone does not induce significant chemical changes. After \SI{60}{\minute} of laser illumination in vacuum, no appreciable change in intensity is observed, even though residual H$_2$O is present in the sample chamber and on the sample. Then, dry oxygen ($<$3 ppm H$_2$O) is introduced in the optical cell and the sample is exposed to radiation for another \SI{60}{\minute}. A slight change in Raman intensity is observed, but the overall signal loss is minor because of the limited amount of H$_2$O available at the sample. Finally, the sample is exposed to ambient air and continually illuminated for another \SI{60}{\minute}. It takes about \SI{15}{\minute} for the intensity to change significantly. This delay is attributed to a relatively slow nucleation and initial growth of degradation by-products for thicker samples. From that point on however, a rapid decline of intensity takes place, losing more than one order of magnitude within less than one hour.  Curve C in \fref{Fig:OxidationVac} shows the Raman intensity from a  \SI{120}{\nano\meter} flake measured in vacuum and then in deoxygenated water vapor. The Raman intensity does not show any appreciable change after 1 hour in each environment. Introduction of oxygen in the chamber at the 120 min mark leads to a rapid decrease in the Raman intensity. These experiments establish that continuous above-gap illumination, water vapor and oxygen alone do not support the photo-oxidation process, but that the simultaneous presence of all three is required.

Although the observed oxidation byproducts are those observed from thermally-activated oxidation (Eqs. (S1a)-(S1c) in the Supplementary Material), the photo-oxidation process observed in this work is not temperature driven. The Raman shifts observed correspond to a temperature variation of at most \SI{43}{\kelvin} at the studied exposure powers (see \S SII in the Supplementary Material for the detailed analysis). This result confirms that the above gap illumination is not contributing significant heat, but rather generating an carrier density $\delta \eta$ in the conduction band.


The results presented thus far suggest that the photo-oxidation involves these three steps: optical excitation of carriers, electron transfer towards aqueous ($aq$) oxygen molecules at the surface of the sample, and spontaneous reaction between aqueous oxygen superoxide anions (O$_{2 (aq)}^{\cdot-}$) with the $p$-doped semiconductor. This process can be modeled using the following equations:
\begin{subequations}
\begin{align}
{\rm GaSe} + h\nu   &\rightarrow {\rm GaSe} ^*\\
{\rm GaSe} ^* + {\rm O}_{2(aq)} &\rightarrow {\rm O}_{2 (aq)}^{\cdot-} + {\rm GaSe} + h^+ \label{CT}\\
{\rm O}_{2 (aq)}^{\cdot-} + {\rm GaSe} + h^+ &\rightarrow \{\rm GaSe\}_{ox}\label{ox},
\end{align}
\end{subequations}
\noindent \noindent where $\rm{GaSe}^*$ represents GaSe with photoexcited electron hole pairs, and $\{\rm GaSe\}_{ox}$ represents GaSe oxidation byproducts. Although the photo-driven dynamics are expected to be quite different from the thermal dynamics discussed in Ref. \onlinecite{Balitskii2004}, our results confirm that the oxidation byproducts are similar.

The charge transfer between the photo-excited carriers and aqueous oxygen (Eq. \eqref{CT}) can be understood in the framework of the Marcus-Gerischer theory, where the rate of charge transfer is described by \cite{Favron2015} \cite{Favron2015} :
\begin{align}
\frac{d [O_{2}^{\cdot-}]}{dt} &\propto \delta\eta[O_2]\exp{\left[-\frac{\left(\chi_{\rm{GaSe}}-E^0_{F,redox}-\lambda\right)^2}{4k_BT\lambda}\right]},\label{MGT}
\end{align}
\noindent where $\delta\eta$ is the conduction band electron density, $[O_{2}]$ and $[O_{2}^{\cdot-}]$ are the concentrations of neutral oxygen and superoxide in the water at the sample surface, $\chi_{\rm{GaSe}}$ is the GaSe electron affinity, $E^0_{F,redox}$ is the Fermi level of the oxygen-water solution and $\lambda$ is the solvent reorganization energy around oxygen molecules. Since GaSe has an electron affinity which positions the lower conduction band in line with the aqueous oxygen acceptor states, the exponential function in Eq. \ref{MGT} enables charge transfer towards the aqueous O$_2$ (see \S SVI in the supplementary material). The generated aqueous superoxide anions (O$_{2 (aq)}^{\cdot-}$) at the surface then rapidly react with the p-doped GaSe (Eq. \eqref{ox}) to produce the observed oxidation byproducts, which will in turn decrease the observed GaSe Raman intensity.

\begin{figure}[htb]%
\includegraphics[width=8.5cm]{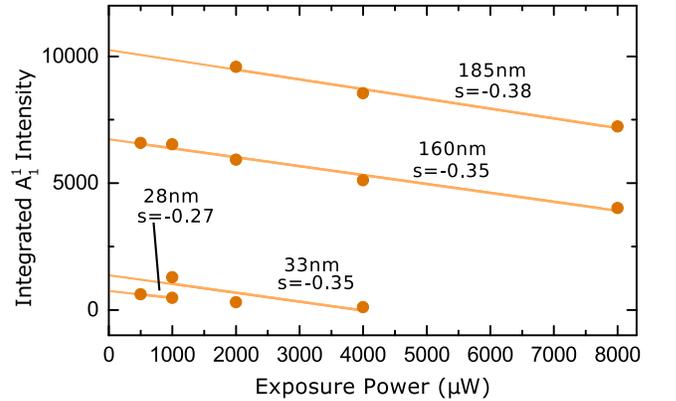}
\caption{Integrated Raman intensity of $A_1^1$ as a function of exposure power for samples of varying thicknesses. Each data point represents a \SI{15}{\minute} exposure at the specified power before a Raman spectrum was measured using a \SI{500}{\mu W} laser excitation. Linear regression slopes $s$ are shown.}
\label{Fig:Power}%
\end{figure}

In the thickness regime studied here, the band gap variation is negligible,\cite{Rybkovskiy2011,Hu2013} and the absorption coefficient is independent of thickness. The rate at which superoxide anions are generated is thus proportional to the exposure power,  $d [O_{2}^{\cdot-}]/dt \propto \delta\eta \propto P_{\rm{exp}}$.  \Fref{Fig:Power} shows the intensity of $A_1^1$ obtained after an exposure of 15 min at the indicated exposure power in ambient air for samples of varying thicknesses. As predicted by Eq. \eqref{MGT}, the observed degradation rate increases linearly with exposure power and is independent of sample thickness, as evidenced from the very similar slopes measured. This confirms the role of the linear dependence of the photo-oxidation mechanism on above gap illumination and further rules out thermal processes. 

 In summary, GaSe exhibits a relatively complex oxidation dynamics compared to other 2D materials. It involves several components : oxygen, humidity, and above-gap illumination, and results in the generation of several oxidation products, Ga$_2$Se$_3$, Ga$_2$O$_3$ and amorphous and crystalline selenium. Luminescence and Raman measurements in ambient conditions are found to severely degrade the optical and electronic characteristics of GaSe, indicating that no safe illumination threshold exists, especially for thin layers of GaSe. Hence, the study of the intrinsic characteristics of 2D GaSe requires a controlled environment at all stages of processing and characterization.

See supplementary material for more detail on the thermal oxidation process and thermal effects, for further analysis of the polarized Raman spectra of the oxide products and of sample photoluminescence, and for additional information regarding the role of water vapor and oxygen in the oxidation process.

This work was made possible by funding from the Fonds de Recherche du Québec-Nature et Technologie (FRQNT) and the Natural Sciences and Engineering Research Council of Canada (NSERC).

\end{document}